\documentclass[aps,prd,showpacs]{revtex4}

\usepackage{psfig}
\usepackage{amsmath,amscd,amssymb,graphicx}
\usepackage[colorlinks=true,urlcolor=green,linkcolor=blue,citecolor=red]{hyperref}
 
\begin{document}

\newcommand{\beq}{\begin{equation}}
\newcommand{\eeq}{\end{equation}}

\draft

\title{Localizing gravity on exotic thick 3-branes 
}

\author{Oscar Castillo-Felisola$^{(1,2)}$, Alejandra Melfo$^{(1)}$,
Nelson Pantoja$^{(1)}$, and Alba Ram{\'\i}rez$^{(1)}$ }

\affiliation{$^{(1)}$ {\it Centro de F\'{\i}sica Fundamental, Universidad de
Los Andes, M\'erida, Venezuela }}

\affiliation{$^{(2)}${\it International Centre for Theoretical Physics,
34100 Trieste, Italy }}

\begin{abstract}

We consider localization of gravity on thick branes with a non trivial
structure. Double walls that generalize the thick Randall-Sundrum
solution, and asymmetric walls that arise from a $Z_2$-symmetric scalar
potential, are considered. We present a new asymmetric
 solution: a
 thick brane
interpolating between two $AdS_5$ spacetimes with different
cosmological constants, which can be derived from a ``fake
supergravity'' superpotential, and show that it is possible to
confine gravity on such branes.

\end{abstract}


\maketitle

\section{Introduction}

As is well known, gravity can be localized on an infinitely thin brane
embedded in a 5-dimensional spacetime \cite{Randall:1999vf}.  It is not
surprising, although perhaps less known, that this result holds true
when the infinitely thin brane is replaced by its regularized version
\cite{Gremm:1999pj}, a thick wall whose distributional thin-wall limit
reduces rigorously to the Randall-Sundrum (RS) spacetime
\cite{Guerrero:2002ki}.  
Thick domain walls are solutions to the coupled Einstein-scalar field
equations with an appropriate discrete symmetry-breaking potential and
boundary conditions that ensure topological stability.

Of course, domain wall solutions long predate the Randall-Sundrum
result, and one can find a number of domain wall spacetimes with
interesting features \cite{g90,Gass:1999gk}. Afterwards, many other
wall solutions have been found, and in some cases shown to also
localize gravity. This, however, is not always
the case.

 In this paper, we address the issue of localizing gravity in
some exotic types of domain walls. First, we present a new method for 
analyzing the spectrum of perturbations on the walls, which greatly
simplifies the calculations when dealing with more involved spacetimes.

The regularized RS spacetime of Ref. \cite{Gremm:1999pj} is a particular
case of the more general thick wall found in \cite{Melfo:2002wd}. In the
general case, these are so-called ``double'' walls, that is, the
energy density is not peaked around a certain value of the bulk
coordinate, but has a double peak instead, representing two parallel
walls, or, more exactly, a wall with some nontrivial internal structure.
 Similar double walls were later found in \cite{Bazeia:2003qt}  and shown to
localize gravity. We show how this holds true also
for the double walls that are a generalization of the regularized RS walls.

An example of even more exotic walls are those lacking a $Z_2$
symmetry, i.e., walls that are not reflection symmetric although the
scalar field potential is $Z_2$ invariant. A particular example of
these walls was found in \cite{Gass:1999gk}, and it is remarkable
that this solution can be found for the same scalar field potential
 than the usual,
reflection-symmetric one. This result, that the same scalar field
configuration and potential can produce two different spacetimes, one
reflection symmetric but dynamic, one asymmetric but static, was found
to be true for a general class of walls in \cite{Guerrero:2002ki}.
 It is therefore
interesting to investigate whether the two types of solutions can
confine gravity in a similar way.  We find below that these type of 
asymmetric walls
cannot confine gravity, even when their dynamic counterparts do.

A perhaps more interesting type of asymmetric brane  is the one that
interpolates between spacetimes with different cosmological constants,
or, equivalently, one where the scalar field interpolates between two
non-degenerate minima of the potential.
Thin brane-world scenarios in which the reflection symmetry along
the extra dimension was broken by gluing two $AdS_5$ spacetimes
with different cosmological constants have been considered in
\cite{Ida:1999ui,Deruelle:2000ge,Stoica:2000ws,Perkins:2000zp}
\footnote{Within
the domain wall context, non-reflection symmetric thin domain
walls between spacetimes with negative and zero cosmological
constant have been discussed previously in \cite{Jensen:1996uv}.}.
Classically, these thin walls should arise as a well defined
distributional limit of smooth field configurations, i.e. thick
walls, a fact that is far from being obvious due to the
nonlinearities of general relativity \cite{Guerrero:2002ki}.
 To our knowledge, however, a thick
wall with these features has not been reported. We find
here one such solution, which in addition can be shown to be derivable from a
superpotential \cite{Behrndt:1999kz,Skenderis:1999mm,DeWolfe:1999cp} in the
context of the so-called fake 
supergravity
\cite{Freedman:2003ax} (for domain wall solutions of the RS type
in supergravity, see \cite{Behrndt:2001km}). Localization of gravity in this wall is then shown
to be possible.

\section{Gravitational perturbations}

We wish to obtain the equations for the perturbations to an
arbitrary solution of the Einstein-scalar field coupled system. In
the usual approach one considers general fluctuations, with or
without gauge fixing, around a background metric of a given form
\cite{DeWolfe:1999cp,Garriga:1999bq,Alonso-Alberca:2000ne,Karch:2000ct}.
Instead, we present here a straightforward generalization of the
well-known procedure to obtain the perturbation equations of an
arbitrary solution to the Einstein equations in vacuum \cite{Wald}
to the case of an arbitrary solution of the Einstein-scalar field
coupled system.

Let $g_{ab}$ and $\phi$ be exact solutions of 
\beq R_{ab}
-\frac{1}{2}g_{ab} R = T_{ab} ; \quad T_{ab} =\nabla_a \phi
\nabla_b\phi - g_{ab} (\frac{1}{2} \nabla^c \phi \nabla_c \phi  +
V(\phi)) ; \quad \nabla_a\nabla^a\phi =\frac{dV}{d\phi}
\label{esf} \eeq where $\nabla_a$ is the derivative in $g_{ab}$.
Now, suppose that a one-parameter family of metrics $
\tilde{g}_{ab}(\lambda)$ and a one-parameter family of scalar
fields $\tilde{\phi}(\lambda)$ exist, such that they satisfy
(\ref{esf}). Suppose also that
\begin{description}
\item[{\em i)}] $\tilde
g_{ab}(\lambda)$ and $\tilde\phi(\lambda)$ depend
differentiably on $\lambda$,
\item[{\em ii)}]
  $\tilde{g}_{ab}(\lambda)|_{\lambda=0} = g_{ab}$ and
  $\tilde{\phi}(\lambda)|_{\lambda=0} = \phi$
\end{description}
We expect that small $\lambda$ corresponds to small
deviations from $g_{ab}$ and $\phi$, hence
\beq
\frac{d}{d\lambda} \tilde{g}_{ab}(\lambda)|_{\lambda=0}= h_{ab} \quad \frac{d}{d\lambda} \tilde{\phi}(\lambda)|_{\lambda=0}= \varphi
\eeq
where $h_{ab}$ and $\varphi$ are the metric and scalar field
perturbations, respectively.
 
Now consider eqs.(\ref{esf}) for $\tilde{g}_{ab}$ written in Ricci
form, for a 5-dimensional spacetime
\beq \tilde{R}_{ab} = \tilde{T}_{ab} -
\frac{1}{3}\tilde{g}_{ab}\tilde{T}; \quad \tilde{T}_{ab}
=\tilde{\nabla}_a\tilde{\phi} \tilde{\nabla}_b\tilde{\phi} -
\tilde{g}_{ab} (\frac{1}{2} \tilde{\nabla}^c \tilde{\phi}\tilde{
\nabla}_c\tilde{ \phi}  + \tilde{V}(\tilde{\phi})) \eeq
 where
$\tilde{T} = \tilde{g}^{ab}\tilde{T}_{ab}$. For the scalar field
$\tilde{\phi}$ we have 
\beq
\tilde{g}^{ab}\tilde{\nabla}_a\tilde{\nabla}_b\tilde{\phi}
=\frac{d\tilde{V}}{d\tilde{\phi}} \eeq

Then, evaluating the $\lambda$-derivatives in $\lambda=0$ of
the above, we have for the linearized Einstein-scalar field
equations for the metric perturbation $h_{ab}$ and the
scalar perturbation $\varphi$ of the exact solutions
$g_{ab}$ and $\phi$:

\begin{eqnarray}
-\frac{1}{2}\nabla^d\nabla_d h_{ab} - \frac{1}{2}\nabla_a\nabla_b
(g^{cd}h_{cd}) + \nabla_{(a}\nabla^ch_{b)c} + R^{c\,\,\, d}_{(ab)} h^{cd} +
R_{(a}^c h_{b)c}= \nonumber \\
\frac{2}{3}h_{ab} V(\phi) + \frac{2}{3}g_{ab}\left[g^{cd} h_{cd}
(\frac{1}{2} \nabla^e\phi\nabla_e\phi + V(\phi))
+\frac{dV}{d\phi}\varphi  \right] + 2
\nabla_{(a}\phi\nabla_{b)}\varphi \label{pert1}
\end{eqnarray}
and 
\beq 
- h^{ab}\nabla_a\nabla_b \phi -\frac{1}{2} g^{ab} g^{cd}
(\nabla_a h_{bd} + \nabla_b h_{ad} - \nabla_d h_{ab}) \nabla_c\phi
+ g^{ab} \nabla_a\nabla_b \varphi - \frac{d^{2}V}{d\phi^{2}}\varphi = 0.
\label{pertesc} \eeq

 A domain wall spacetime is a solution of
eq.(\ref{esf}) with plane-parallel symmetry, with the scalar field
depending only on the bulk coordinate, $\phi=\phi(\xi)$. Choosing
the axial gauge 
\beq h_{\xi a} =0, \label{axial} \eeq 
the
transverse traceless (TT) part of $h_{ab}$ decouples from the
scalar field fluctuations, as can be seen from (\ref{pertesc}).
Therefore, working in this gauge, we can set the field
fluctuations $\varphi$ to zero. For the TT modes, eq.(\ref{pert1})
reduces to 
\beq -\frac{1}{2}\nabla^d\nabla_d h_{ab} + R^{c\,\,\,
d}_{(ab)} h_{cd} + R_{(a}^c h_{b)c} = \frac{2}{3}h_{ab} V(\phi)
\label{pert} \eeq
 which describes linearized gravity in the
transverse and traceless sector. We shall use this equation to
find the TT modes for different domain wall spacetimes.

\section{Double walls}

\begin{figure}
\centerline{\psfig{figure=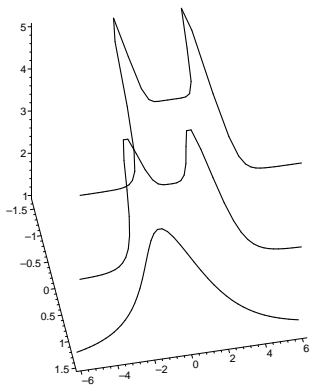,height=5cm}}
\caption{Energy density as a function of $\xi$ for double  walls with $s=1,3,5$.\label{2rho}}
\end{figure}

In \cite{Melfo:2002wd}, a solution to eq.(\ref{esf}) was found for the
static spacetime of a domain wall with an internal structure. This
so-called double wall solution is given by
\begin{equation}
g_{ab}= e^{2 A(\xi)}( \eta_{ab} + d\xi_{a}d\xi_{b}), \qquad A(\xi)=
-\frac{1}{2s}\ln( 1 + (\alpha \xi )^{2s}),
 \label{doublemetric}
\end{equation}
where $\eta_{ab}$ is the 4-dimensional Minkowski metric, and
 \begin{equation} \phi = \phi_0\tan^{-1}(\alpha^s\xi^s) ,  \quad
\phi_0=\frac{\sqrt{3(2s-1)}}{s} ,
\end{equation}
with a potential
\begin{equation}
V(\phi) + \Lambda =3
\alpha^2\sin(\phi/\phi_0)^{2-2/s}\left[\frac{2 s + 3}{2}
\cos^2(\phi/\phi_0) - 2 \right]. \label{vstat}\end{equation} For
$s=1$ it reduces to a well-known domain wall solution which is a
regularized version of the RS thin brane
\cite{Gremm:1999pj,Guerrero:2002ki}, while for odd  $s>1$ the energy density is
peaked around two values, as can be seen in Fig. \ref{2rho}. These walls
interpolate between anti-de Sitter asymptotic vacua with
$\Lambda=-6\alpha^2$ \cite{Melfo:2002wd}. Similar solutions were also
considered in \cite{Bazeia:2003qt}. We shall show that these double walls
can confine gravity.

\begin{figure}
\centerline{\psfig{figure=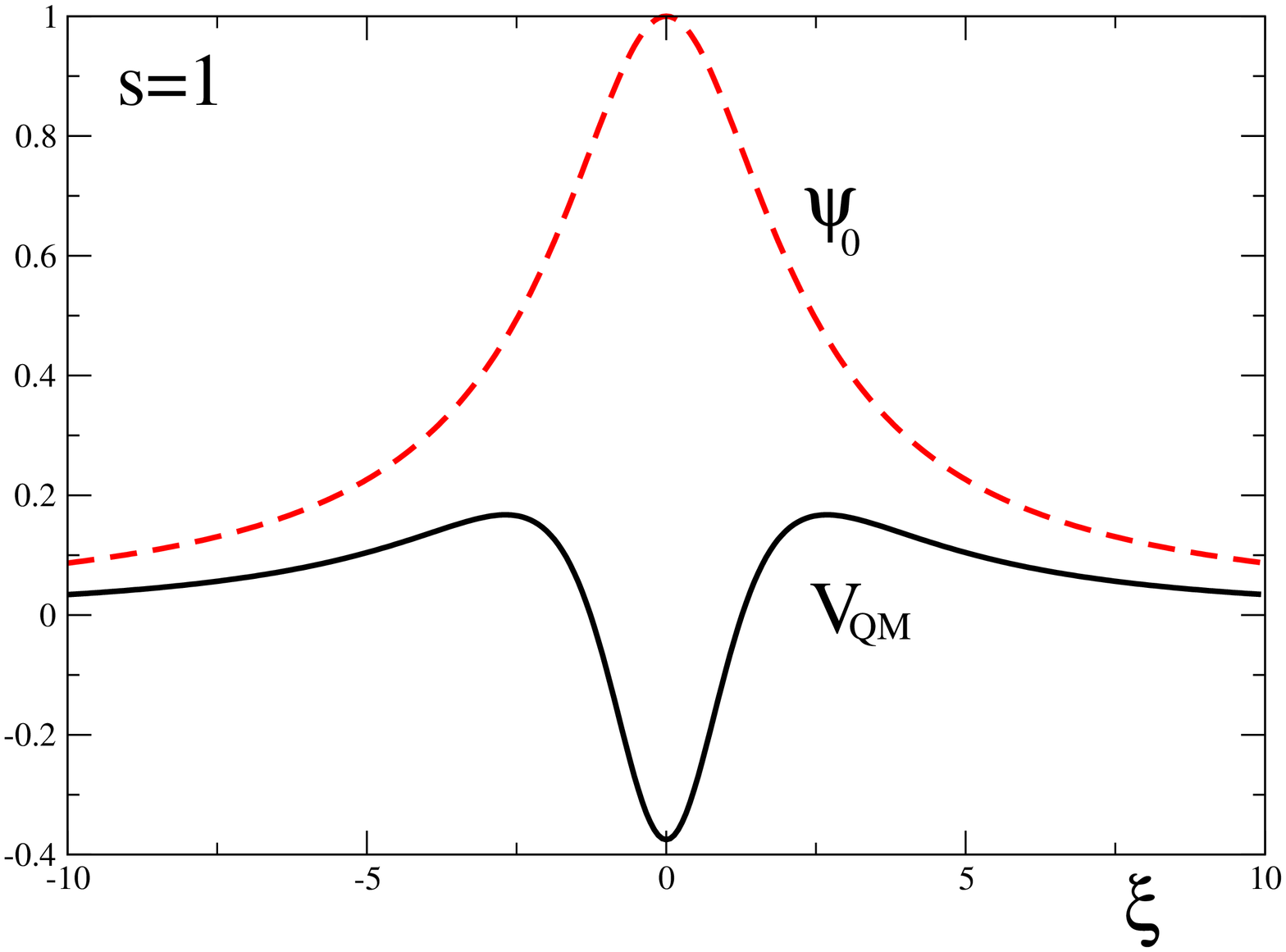,height=5.5cm,angle=-90}\psfig{figure=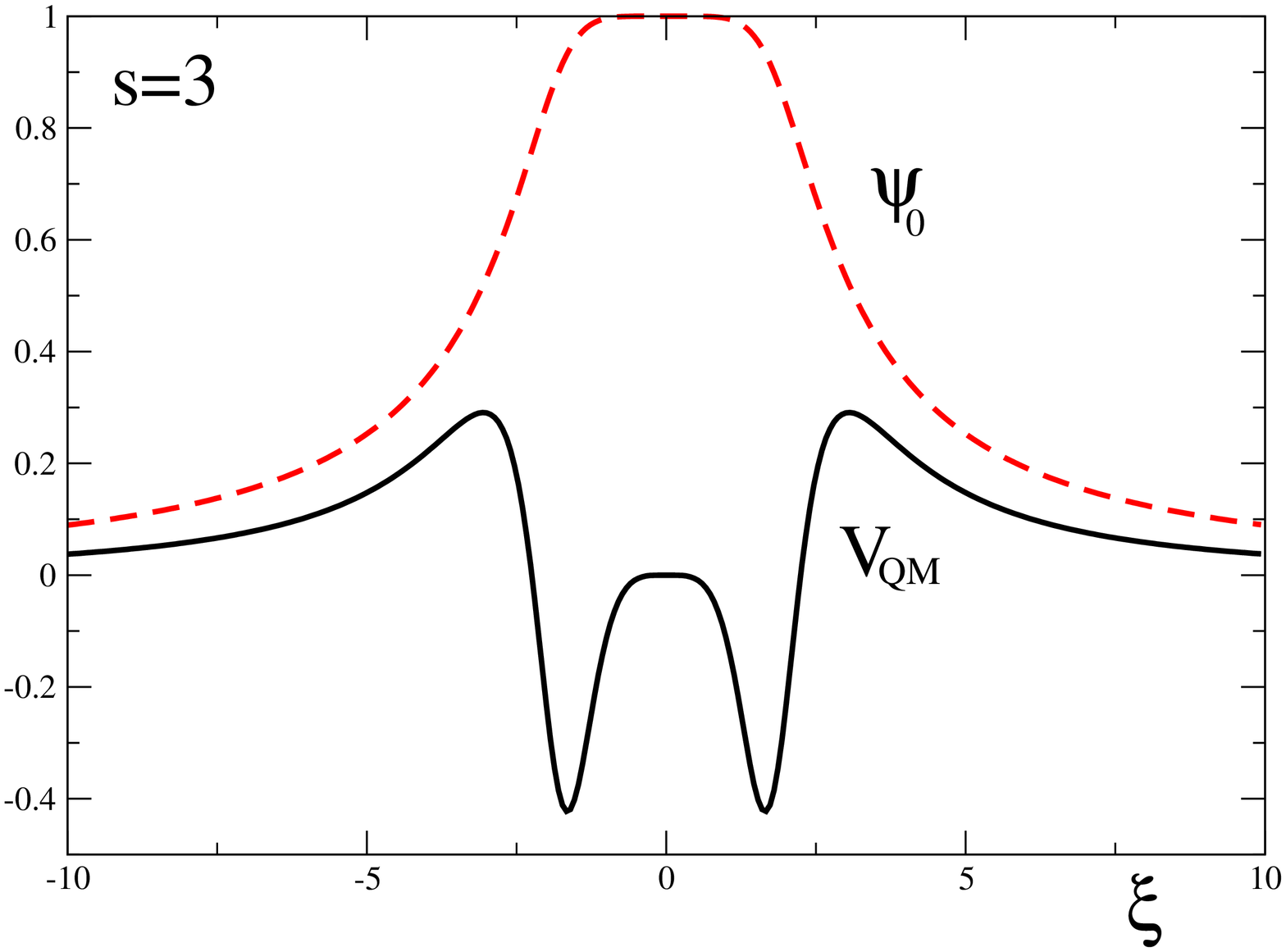,height=5.5cm,angle=-90}
}
\caption{$V_{QM}$ and the zero mass mode for 
 walls with $s=1$ and $s=3$\label{2qm}.}
\end{figure}

We write the metric fluctuations conveniently as
 \beq h_{\mu\nu}= e^{\imath p\cdot  x} e^{A(\xi)/2} \psi_{\mu\nu}(\xi) \label{pert2}\eeq where
$\mu, \nu = 0...3$. From eq.(\ref{pert}) we have
 \beq
(-\partial_\xi^2 + V_{QM} ) \psi_{\mu\nu} = m^2 \psi_{\mu\nu} \eeq
where \beq V_{QM}(\xi) = \frac{3}{4\xi} \frac{5 (\alpha\xi)^{4s} +
2 (\alpha\xi)^{2s} - 4 s (\alpha\xi)^{2s}}{(1 +
(\alpha\xi)^{2s})^2} \eeq
 The zero modes for each $s$ are easily
found to be ($\mu$, $\nu$ indices omitted)
\beq \psi_0 = N \left[ (1 + (\alpha\xi)^{2})
\sum_{m=0}^{s-1} (- \alpha^2\xi^2)^m\right]^{-3/4s} 
 \eeq 
which are
indeed the lowest mass ones. 
There is one discrete bound state at the threshold
and one can show the existence of a continuum of states that
asymptote to plane waves as $\xi\rightarrow\pm\infty$, as follows form
the fact that $V_{QM} \to 0$ asymptotically. The
potential $V_{QM}$ and the zero mode are plotted in Fig. \ref{2qm} for $s$ odd.

\section{Asymmetric vs. Dynamic walls}

In \cite{Melfo:2002wd}, it was shown that for a general class of scalar field
potentials, it is possible to obtain two different domain wall
solutions: one representing a dynamic spacetime with a $Z_2$ symmetry
on the wall's plane, and another lacking this symmetry, and with a
static metric. Both solutions have  essentially the same scalar field
configuration. We wish to address in this section the issue of whether
these configurations can localize gravity.

The most general 5-dimensional time-dependent metric representing
the spacetime of a static domain wall can be written as
\begin{equation}
g_{ab}= e^{2 A(\xi)}\left[ -dt_{a}dt_{b}  + e^{2 \beta t} dx^i_{a}
dx^{i}_{b} + d\xi_{a}d\xi_{b} \right],
 \label{dynamicmetric}\end{equation}
where $x^i$ ($i=1,2,3 $) are spatial coordinates on the wall. When 
\beq A(\xi)
= -\delta \ln\left( \cosh ({\beta\xi}/{\delta})\right), \eeq
eq.(\ref{dynamicmetric}) is a solution of (\ref{esf}) with
\begin{eqnarray} \phi(\xi) &=&  \phi_0
\tan^{-1}[\sinh(\beta\xi/\delta)]  , \quad \phi_0 =\sqrt{\frac{3
a^2\beta^2}{4a}}\frac{ \sqrt{\delta(1-\delta)}}{\beta},
\label{phicosh} \\
 V(\phi) &=& \frac{ [1+\delta(a-1)]}{\delta}
\frac{3 a^2 \beta^2}{8 a} [\cos(\phi/\phi_0)]^{2(1-\delta)}
\label{vecosh} \end{eqnarray}
 for $a=4$.  In \cite{Guerrero:2002ki}, it was
shown  that this solution has a well-defined thin wall limit when
$\delta\to 0$, such that the metric and all the curvature tensor
fields make sense as distributions. The analysis of gravitational
fluctuations around this geometry, parameterized in a slightly
different manner and for the case of a finite width, where studied
in \cite{Wang:2002pk}.

From eq.(\ref{pert},\ref{pert2}) we find that the spectrum
of perturbations consists of a zero mode 
\beq \psi_o = N \left[ \cosh
\left( {\beta \xi}/{\delta}\right ) \right ]^{-3\delta/2} \eeq
 and
a set of continuous modes separated by a mass gap given by
$\frac{9}{4}\beta^2$, as in \cite{Wang:2002pk}. Interestingly enough, the
thin wall limit for the gravitational perturbations can also be
obtained in the sense of distributions. We find
\begin{equation}
\lim_{\delta\rightarrow 0}\psi_0= \frac{2}{3\beta}\exp({-3\beta|\xi|/2}).
\end{equation}
On the other hand
\begin{equation}
\lim_{\delta\rightarrow 0}V_{QM}= \frac{9}{4}\beta^2 -
3\beta\delta(\xi)
\end{equation}
which, besides the $\delta$-term responsible for the unique bound
state, clearly shows the mass gap whose existence is a generic
property of the de Sitter branes
\cite{Garriga:1999bq,Alonso-Alberca:2000ne,Karch:2000ct,Wang:2002pk}.

Now, as we argued above,  the metric
(\ref{dynamicmetric}) is not the only solution for a scalar field
potential of the form (\ref{vecosh}).  There is a corresponding 
static solution with a static  metric
that is not reflection symmetric on the wall's plane, 
still given by eq.(\ref{vecosh}), with
$a=1$, but with a static metric 
\begin{equation}
g_{ab}= e^{A(\xi)/2 - 3\beta \xi/2} \left(-dt_{a}dt_{b} +
e^{2\beta\xi} dx^i_{a} dx^{i}_{b} \right) + e^{2A(\xi)}d\xi_{a}d\xi_{b}.
\label{metricA}
\end{equation}
It should be stressed that these asymmetric thick branes arise as
solutions to the Einstein-scalar field equations with a $Z_2$
symmetric potential for which $V(\phi(-\infty))= V(\phi(\infty))=0$.
 The thin-wall limit of this spacetime can also
be taken rigorously \cite{Melfo:2002wd,Pantoja:2003zr}. At one side of the wall, the
spacetime is asymptotically Minkowski, while on the other side it
tends to the Taub \cite{Taub:1950ez} spacetime.\footnote{Static thin
domain walls without reflection symmetry embedded in spacetimes
with vanishing cosmological constant have been considered
previously in \cite{Tomita:1985bg}.}

Most of the thin brane-world scenarios assume that the extra
dimension is $Z_2$-symmetric, although asymmetric brane-world
scenarios can also be considered. We now show that the asymmetric
domain wall spacetime with metric (\ref{metricA}) cannot confine
gravity. Let us consider the spectrum of linearized gravity
fluctuations around (\ref{metricA}). For this background geometry,
from (\ref{pert}) and in the axial gauge (\ref{axial}), it follows
that the only non-zero TT modes are $h_{ij}$ with $i,j= 1,2,3$.
Next, for these modes, with the rescaling
\begin{equation}
h_{ij}= e^{(A(\xi) + \beta\xi)/2}\psi_{ij}(\xi)
\end{equation}
we obtain
\begin{equation}
\left( \partial^2_{\xi} + e^{2 A} \Box^{(4)} \right) \psi_{ij} =
0,\label{inviable}
\end{equation}
where $\Box^{(4)}$ is the $(3+1)$-dimensional covariant
d'Alembertian
\begin{equation}
\Box^{(4)}= e^{-(A + \beta\xi)/2}\left(
-e^{2\beta\xi}\partial^2_t +
\partial^2_x + \partial^2_y +
\partial^2_z\right).\label{fakedalembertian}
\end{equation}
Now, we find that the linearized equation of motion for tensor
fluctuations (\ref{inviable}) can not be rewritten as a
Schr\"{o}dinger equation, (\ref{fakedalembertian}) depends
explicitly of the additional coordinate $\xi$ through $A(\xi)$
and all the modes propagate in the bulk. Hence, a normalizable zero mode
cannot be found, and the usual method of demonstrating localization of gravity
cannot be applied to this spacetime.
Notice that (\ref{metricA}) does not belong to the
general class of metrics considered for example in \cite{Csaki:2000fc}. 
One could try to add more ingredients to the
theory (e.g. additional branes \cite{Gregory:2000jc}, mass terms for the matter
fields \cite{Dubovsky:2000am},  curvature
terms on the brane \cite{Dvali:2000hr}, etc.),
 we do not pursue this direction here.

\section{An asymmetric thick brane-world scenario}

Let us now consider an asymmetric thick domain wall spacetime
where the $C^\infty$ metric tensor is
\beq 
g_{ab}= e^{2
A(\xi)}\left( -dt_{a}dt_{b}  + dx^i_{a} dx^{i}_{b} \right) + d\xi_{a}d\xi_{b},
\label{albametric} \eeq
 where 
\beq A(\xi)=-\frac{1}{12}\left[-\alpha \xi + \delta \exp(-2
\exp(-\beta\xi/\delta )) - \delta\,
{\rm Ei}\left(-2\exp(-\beta\xi/\delta)\right) \right], \label{albamu}
\eeq
 with Ei the exponential integral given by
 \beq {\rm Ei}(u)
\equiv -\int_{-u}^\infty d\tau \frac{e^{-\tau}}{\tau} \eeq
 and
where $\alpha,\beta,\delta$ are real constants with $\delta>0$.
This represents a three-parameter family of plane symmetric static
domain wall spacetimes without reflection symmetry along the
direction perpendicular to the wall, being asymptotically ({\it
i.e.} far away from the wall) $AdS_5$ with a cosmological constant
$-\alpha^2/48$ for $\xi < 0$ and $-(\beta - \alpha)^2/48$ for
$\xi > 0$.

The metric (\ref{albametric}) is a solution to the coupled
Einstein-scalar field equations (\ref{esf}) with 
\beq \phi(\xi)= \sqrt{\delta} \exp(- \exp(-\beta\xi/\delta ))
\label{albaphi} 
\eeq
and 
\beq V(\phi)= \frac{1}{8}\left[\frac{\beta^2}{\delta^2}
\phi^2\ln^2\left(\frac{\phi^2}{\delta}\right)
- \frac{4}{12}\left(\frac{\beta}{\delta}\phi^2 \left(1 -
\ln\left(\frac{\phi^2}{\delta}\right) \right) - \alpha
\right)^2\right], \label{albapot} \eeq 
where $\phi$ interpolates
between the two non degenerate minima of $V(\phi)$, $\phi_0 = 0$ and
$\phi_1 = \sqrt{\delta}$ (see Fig. 3a), and where $\delta$
plays the role of the wall's thickness. Following \cite{Guerrero:2002ki},
the distributional $\delta\to 0$ thin wall limit of this geometry
can be obtained and we find 
\beq \lim_{\delta\to 0} e^{2A(\xi)}
=e^{{\alpha\xi}/6}\,\Theta(-\xi) + e^{-(\beta -
    \alpha)\xi/6}\,\Theta(\xi),\label{twl}\eeq 
where $\Theta$ is the
Heaviside distribution. This clearly shows that this spacetime
behaves asymptotically \cite{Pantoja:2003zr} as an $AdS$ spacetime
with different cosmological constants at either side of the wall.
Furthermore, it also shows that the emergence of two separate
$AdS_5$ patches with different cosmological constants can arise
(rigorously) from the thin wall limit of a thick domain wall
interpolating between non degenerated minima of the scalar
potential.

Remarkably, using the first order formalism of
\cite{Behrndt:1999kz,Skenderis:1999mm,DeWolfe:1999cp},
(\ref{albametric}-\ref{albamu}) and  (\ref{albaphi}-\ref{albapot})
can be obtained from a single function, the superpotential
$W(\phi)$ given by  
\beq W(\phi) = \frac{1}{24} \left[
\beta\frac{\phi^2}{\delta}  \left(1 -
\ln\left(\frac{\phi^2}{\delta}\right) \right) - \alpha \right]
\label{albaspot} \eeq 
 whose critical points are the asymptotic
values of $\phi$ as given by (\ref{albaphi}) (see Fig. 3a). 
Now, critical points
of $W$ are also critical points of $V$ and in the context of
supergravity theories the critical points of $W$ yield stable
$AdS$ vacua \cite{Skenderis:1999mm}. Interestingly, the vacuum
structure of the present model is described by a fake
superpotential which resemble the true ones that appear in low
energy global supersymmetric effective field theories.

Next, let us examine the stability of this domain wall spacetime
as a thick brane-world scenario. In the axial gauge (\ref{axial}),
by making the change of variables $d\chi\equiv e^{-A}d\xi$ and
writing the metric fluctuations as 
\beq h_{\mu\nu} = e^{\imath p\cdot
x} e^{-3 A(\chi)/2} \psi_{\mu\nu}(\chi) \eeq
 where $\mu, \nu = 0...3$, we find from eq.(\ref{pert}) 
\beq
(-\frac{1}{2}\partial_\chi^2 + V_{QM} ) \psi_{\mu\nu}(\chi) = m^2
\psi_{\mu\nu}(\chi),\label{schro} \eeq where \beq V_{QM}= \frac{1}{2}\left(\frac{9}{4} A'(\chi)^2 + \frac{3}{2}
A''(\chi)\right) \label{VQM}\eeq
 For the zero mass modes this is
integrated as usual \beq \psi_0(\chi)= N_0 e^{3 A(\chi)/2} \eeq
which is a normalizable mode if $\beta
> \alpha > 0$. The zero mode and $V_{QM}$ are given in Fig. 3b.

\begin{figure}
\centerline{\psfig{figure=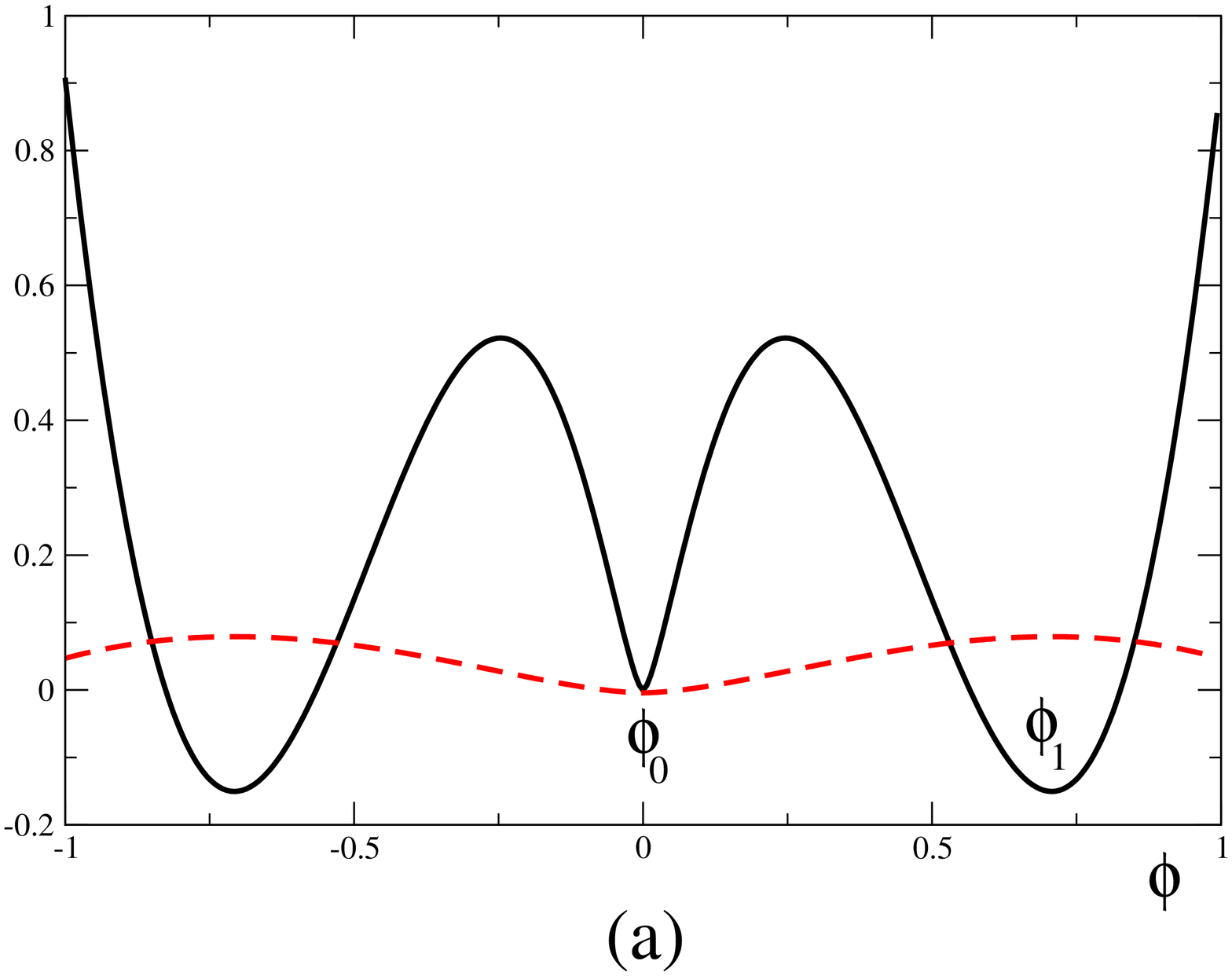,height=5.5cm,angle=-90}\hspace{1cm}\psfig{figure=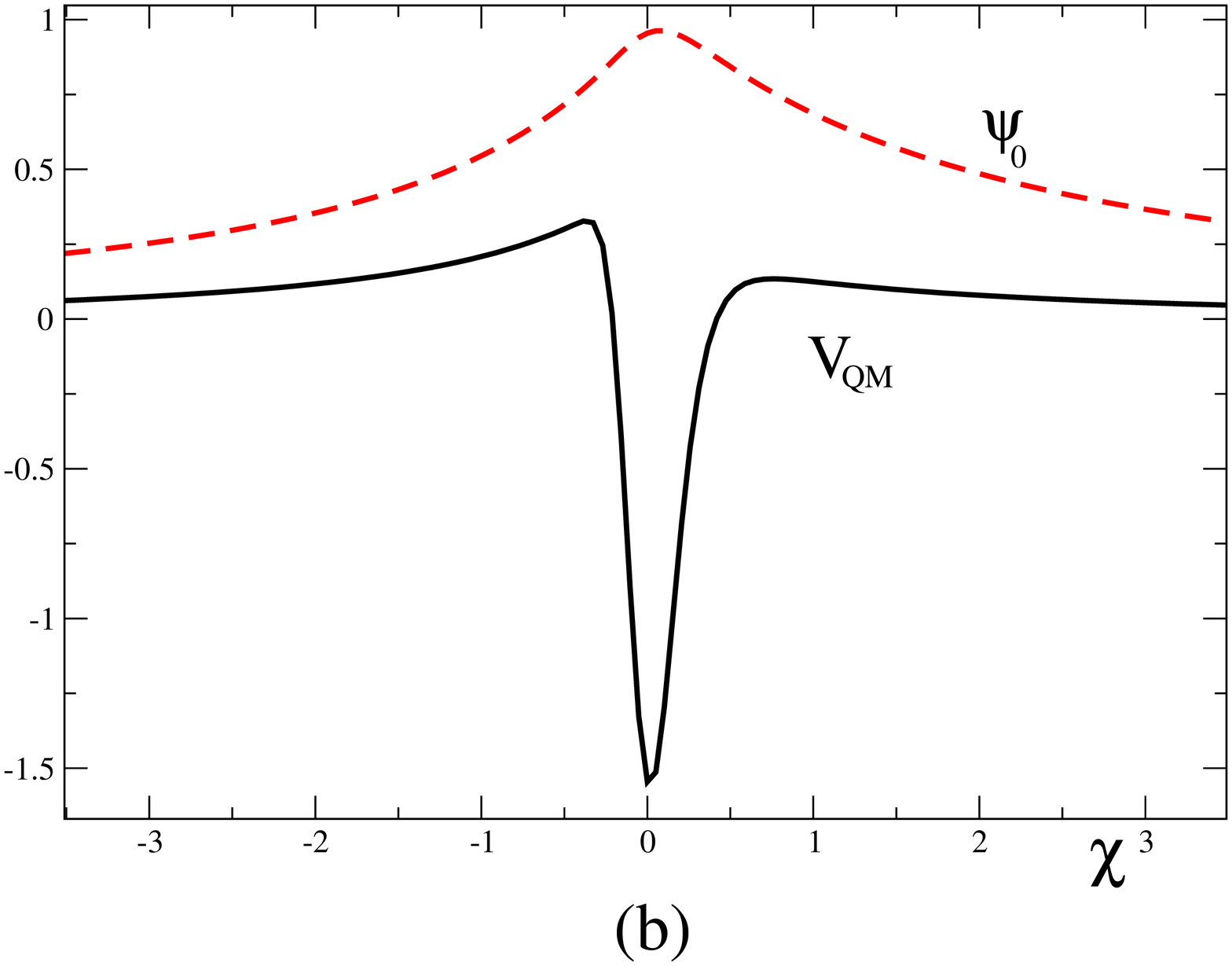,height=5.5cm,angle=-90}}
\caption{Figure (a) shows the scalar field potential (continuous line)
and superpotential (dashed line) for the asymmetric brane (30-34), with
$\beta>\alpha>0$. Figure (b) gives $V_{QM} $  and the
(arbitrarily normalized) zero mode  for the same values of the parameters.}
\end{figure}

 We were not able to find exact solutions
for $m^2>0$. Nevertheless, in the thin-wall limit we can find the
exact massive modes. From (\ref{VQM}) we find 
\beq \lim_{\delta\to 0} V_{QM}=-\frac{3}{48}\beta\delta (\chi) +
\frac{15}{8}\frac{\alpha^2}{(12-\alpha\chi)^2}\Theta (-\chi) +
\frac{15}{8}\frac{(\beta-\alpha)^2}{(12+(\beta-\alpha)\chi)^2}\Theta
(\chi).\label{asymQM}\eeq 
Hence, we expect a single normalizable
bound state mode at the threshold and a continuum of Kaluza-Klein
states for all possible $m^2>0$. In fact, in this limit, the
zero mass solution of (\ref{schro}) is given by 
\beq
\lim_{\delta\to 0}\psi_0(\chi)\sim (1-k_-\chi)^{-3/2}\Theta
(-\chi) + (1+k_+\chi)^{-3/2}\Theta (\chi),
\label{zeromode}
\eeq
 where $k_- \equiv
\alpha /12$ and $k_+ \equiv (\beta - \alpha )/12$, while the
massive ones are given by
\begin{eqnarray} \lim_{\delta\to 0}\psi_m (\chi) &
\sim& (k^{-1}_- -\chi)^{1/2}\left[
Y_2(m (k^{-1}_- -\chi)) +  \frac{ 8 k_{-}^{2}}{ \pi m^{2}} C_- J_2(m (k^{-1}_-
-\chi))\right]
\Theta(-\chi)\nonumber \\ & &+\,(k^{-1}_+  + \chi)^{1/2}\left[ Y_2(m
(k^{-1}_+  + \chi)) +  \frac{8 k_{+}^{2}}{\pi m^{2}} C_+ J_2(m (k^{-1}_+
+\chi))\right] \Theta(\,
\chi), \label{asimmass}
\end{eqnarray}
 where $Y_2$ and $J_2$ are
the Bessel functions of order $2$. In (\ref{asimmass}), $C_-$ and
$C_+$ are constants such that, as follows from
(\ref{schro},\ref{asymQM}), $\psi_m$ be continuous and its
derivative discontinuous at $\chi=0$. We find, for small $m/k_{\pm}$
\beq
C_{+} = \left(1 + \frac{k_{+}}{k_{-}}\right)^{-1}  \left[ 1 +
\frac{1}{2}\left( \frac{k_{+}}{k_{-}} -\sqrt{\frac{k_{+}}{k_{-}}}\right)
+ 4 \frac{k_{+}^{2}}{m^{2}} \left( 1 - \left(\frac{k_{-}}{k_{+}}\right)^{3/2}\right)
\right] 
\eeq
and $C_{-}$ is given by interchanging $k_{+}$ and $k_{-}$ above.  As
expected, the 
massive modes asymptote to plane waves. 
For $\beta=2\alpha$ we
have $k_+=k_-$ and we recover the original $Z_2$-symmetric
RS scenario.

 Now we can calculate the gravitational 
potential  between two particles $(m_{1},m_{2})$ induced by the massless and 
the massive modes on the brane in this approximation.  In order to compare
with the $Z_{2}$ case, consider
the nearly symmetric scenario $k_- \sim k_+$. To first order in   $(k_-/k_+  -
1)$, we obtain from (\ref{zeromode}) and (\ref{asimmass})
\beq
V(r) \sim G_{N} \frac{m_{1} m_{2}}{r} +
\int_{0}^{\infty} dm \frac{G_{N}}{k_{e}} \frac{m_{1} m_{2}
e^{-mr}}{r}\frac{m}{k_{e}} \left[1 + 3\left( \frac{k_{-}}{k_{+}} - 1\right
)\right]
\eeq
 where we have defined
 \beq
 k_{e}^{-1} = \frac{k_{+}^{-1} + k_{-}^{-1}}{2}
 \eeq
 and then as usual 
 \beq
 G_{N} = G_{5} k_{e}
 \eeq
with $G_{5}$ the gravitational constant in 5 dimensions.

\section{Summary and Outlook}

We have explored the possibility of localizing gravity in some 
exotic thick brane spacetimes, using a technique that allows
straightforward calculation of the perturbation equations.

First, we studied  the double-wall spacetime which is a generalization
of the regularized (thick) RS spacetime, in the sense that it
reproduces the latter for a particular value of a discrete
parameter. We have shown that there is always a normalizable
zero mode, and that the quantum-mechanical potential for the modes is
such that there is no gap between the massless and massive modes for
any value of the parameter.

 Next, we consider asymmetric walls arising from a $Z_2$ symmetric
scalar field potential, where the asymmetry is induced by the boundary
conditions at spatial infinity, i.e., it is manifest on the spacetime
metric only. These walls share the scalar field configuration
and potential with the more familiar symmetric and dynamic walls, which
have been shown elsewhere to confine gravity. However, these feature
is shown to be lost in the asymmetric ones. 

Finally, we turn to what we consider the central result of this
paper. We have shown that there exist asymmetric thick brane solutions 
interpolating between two non-degenerate minima of a $Z_2$ symmetric
potential. These branes tends  asymptotically to 
 $AdS_5$ spacetimes with different
cosmological constants at each side, and can be derived from a fake
supergravity superpotential. We have found the zero modes for the
metric fluctuations, and shown that gravity can be localized on this
brane. Furthermore, their thin-wall limit is well defined, and we
find the massive modes on this limit. When the two
cosmological constants at either side of the wall coincide, the usual RS
result is recovered.

\acknowledgments 
We wish to thank Daniel Morales for fruitful
discussions. This work was supported by FONACIT (Project
No. S1-2000000820). O.C-F. wishes to thank ICTP for hospitality during
the completion of this work.


\providecommand{\href}[2]{#2}\begingroup\raggedright\endgroup

\end{document}